\documentclass[aps,prl,reprint,groupedaddress]{revtex4-1}

\usepackage[pdftex]{graphicx}
\usepackage{graphics}
\usepackage{ upgreek }
\usepackage{ textcomp }
\usepackage{ tabularx }
\begin{document}

%\preprint{}

\title{Ultra-low carrier concentration and surface dominant transport in Sb-doped Bi$_2$Se$_3$ topological insulator nanoribbons}

\author{Seung Sae Hong,$^1$ Judy J. Cha,$^2$ Desheng Kong,$^2$ and Yi Cui$^2$  }
\affiliation{ $^1$Department of Applied Physics}
\affiliation{ $^2$Department of Materials Science and Engineering, Stanford University, Stanford, California 94305, USA}


\date{\today}

\begin{abstract}
A topological insulator is a new state of matter, possessing gapless spin-locking surface states across the bulk band gap which has created new opportunities from novel electronics to energy conversion. However, the large concentration of bulk residual carriers has been a major challenge for revealing the property of the topological surface state via electron transport measurement. Here we report surface state dominated transport in Sb-doped Bi$_2$Se$_3$ nanoribbons with very low bulk electron concentrations. In the nanoribbons with sub-10nm thickness protected by a ZnO layer, we demonstrate complete control of their top and bottom surfaces near the Dirac point, achieving the lowest carrier concentration of 2\texttimes 10$^{11}$cm$^{-2}$ reported in three-dimensional (3D) topological insulators. The Sb-doped Bi$_2$Se$_3$ nanostructures provide an attractive materials platform to study fundamental physics in topological insulators, as well as future applications.

\end{abstract}

%\pacs{}
%\keywords{}

\maketitle

%\section{}
%\subsection{}
%\subsubsection{}

The exotic electronic properties of the surface state, due to its spin-momentum locked Dirac cone in the electronic band structure, define a topological insulator as a new class of quantum matter\cite{Moore2010,Qi2010,Hasan2010,Fu2007,Konig2007}. Moreover, it is predicted to offer exciting physics in condensed matter systems, such as elusive quasi-particles, spin transport, and fault tolerant quantum information processing\cite{Qi2010,Hasan2010,Fu2007,Konig2007,Qi2009,Seradjeh2009,Fu2008}. Bismuth selenide (Bi$_2$Se$_3$) and its relative compounds are one of the most promising candidates to realize the ideal three-dimensional topological insulator due to their large bulk band gap and simple surface band structure\cite{Zhang2009}. There have been significant advances to probe the surface state in these materials by various methods, such as angle-resolved photoemission spectroscopy (ARPES)\cite{Xia2009, Chen2009, Zhang2010} and scanning tunneling microscopy (STM)\cite{Roushan2009, Alpichshev2010, Hanaguri2010}. Transport measurements in bulk crystals have demonstrated the existence of these surface states as well\cite{Analytis2010, Qu2010, Taskin2011}.

\begin{figure}
\centering
\includegraphics[width=65mm]{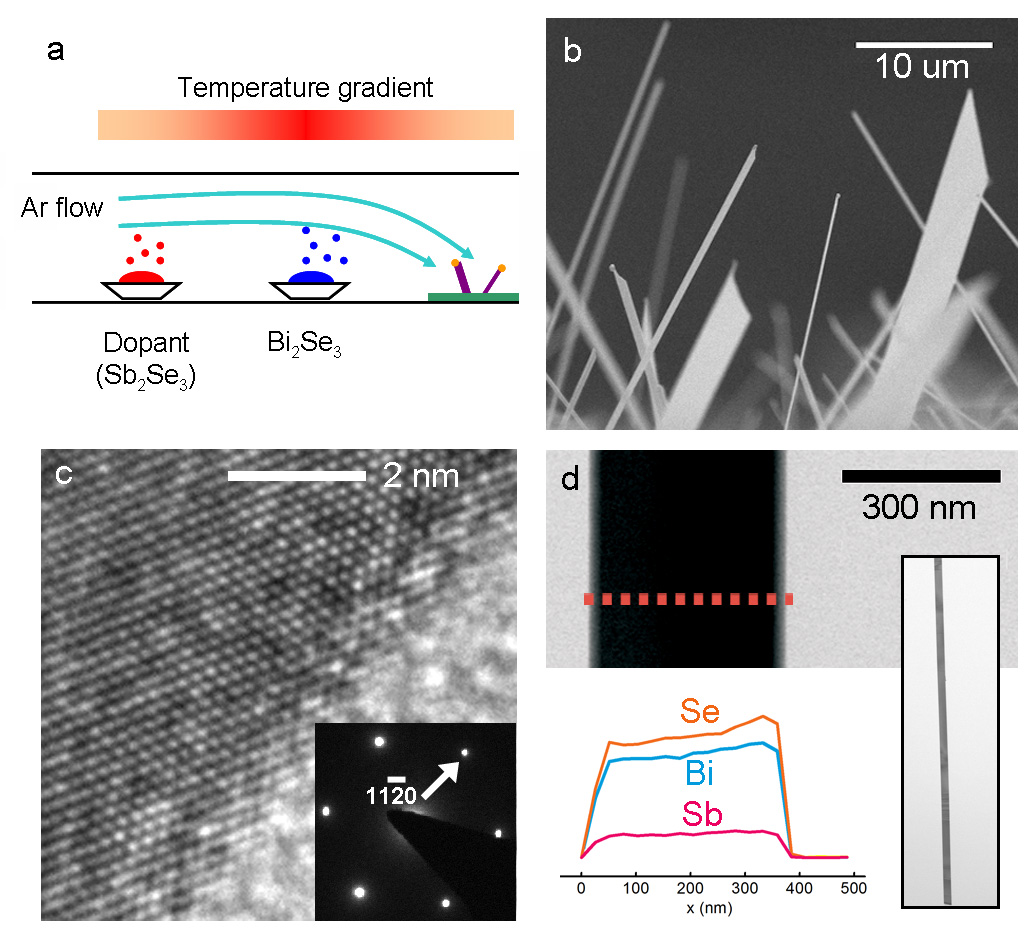}
\caption{\textbf{a}, A schematic of vapor-liquid-solid growth of Sb-doped Bi$_2$Se$_3$ nanoribbons. By shifting the dopant source location along the temperature gradient in the tube furnace, the relative vapor pressure of two sources and the incorporated dopant level are controlled. \textbf{b}, A scanning electron microscopy (SEM) image of as-grown nanoribbons. \textbf{c}, A high resolution TEM image of a nanoribbon and its diffraction pattern (inset). \textbf{d}, A low magnification TEM image of a nanoribbon (top) and its elemental line profile by EDX (bottom), indicating the homogeneous Sb dopant distribution. Inset image shows the entire morphology of the nanoribbon (17$\mu$m long).}
\end{figure}

Nanoscale topological insulator devices offer a unique opportunity to study the surface state as the surface to volume ratio is increased to manifest the surface effect\cite{Peng2009, Xiu2011}. Moreover, on mesoscopic length scale, one can study transport reflecting the fundamental nature of carriers, as shown in previous cases like graphene\cite{Zhang2005, Novoselov2005}. However, in the case of topological insulators, material imperfection issues in the bulk blur surface state signatures and limit further in-depth transport studies. One of the critical obstacles is the dominant bulk electrons outnumbering surface state electrons\cite{Checkelsky2010, Steinberg2011, Kim2011, Kong2011}. Moreover, the material is very sensitive to environmental contamination as sample degradation by environmental exposure has been observed in a few studies\cite{Analytis2010, Kong2011ACS}. Here, we report Sb doping for Bi$_2$Se$_3$ nanoribbons to suppress the bulk conductivity and achieve surface state dominant transport in nanodevices. Transport experiments confirm the significant suppression of the bulk contribution without degradation of electron mobility. In addition, we deposit an oxide layer on top of the nanoribbons as a protective layer, which enables ultralow carrier concentration controllable near the Dirac point and helps realize surface dominant transport in the topological insulator nanostructures.

To suppress the bulk conductivity in topological insulator nanoribbons, we synthesize Bi$_2$Se$_3$ nanoribbons with Sb doping. Sb is known as an effective compensation dopant to reduce bulk electron density to 10$^{17}$cm$^{-3}$ and does not destroy the topological surface state\cite{Analytis2010, Zhang2010APL}. Bi$_2$Se$_3$ nanoribbons are synthesized via vapor-liquid-solid growth mechanism using gold particles as catalysts\cite{Peng2009,  Kong2010}, and Sb vapor is introduced from a Sb$_2$Se$_3$ source material placed at the lower temperature zone (Fig. 1a). As-grown ribbons are 100-300nm thick, their widths vary from 200nm to several micrometers, and their lengths are up to tens of micrometers (Fig. 1b). Sb-doped Bi$_2$Se$_3$ nanoribbons are in a single crystalline rhombohedral phase, which is the same as undoped Bi$_2$Se$_3$ nanoribbons. The distribution of Sb dopants appears to be spatially uniform in the nanoribbon, as confirmed by energy-dispersive X-ray spectroscopy (EDX) (Fig. 1c-1d).

\begin{table}%[H] add [H] placement to break table across pages
\caption{Transport parameters of nanoribbon samples of different Sb concentrations (T=2K)\label{}}
\begin{ruledtabular}
\begin{tabular}{c c c c c}
Sample & Sb & R$_{sheet}$ & R$_{Hall}$ & n$_{Hall}$ \\ 
number & doping & ($\Omega $)  & ($\Omega $) & (10${}^{12}$cm${}^{-2}$) \\ \hline
B1 & 0.0\% & 43 & -7.4 & 84.5 \\ 
B2 & $<$2\% & 177 & -14.7 & 42.5 \\ 
S1 & 3.0\% & 422 & -39.9 & 15.7 \\  
S2 & 5.5\% & 975 & -51 & 12.3 \\  
S3 & 7.0\% & 1350 & -63.1 & 9.9 \\ 
\end{tabular}
\end{ruledtabular}
\end{table}

The Sb doping concentration in Bi$_2$Se$_3$ nanoribbons can be tuned systematically by controlling the temperature of the Sb$_2$Se$_3$ source material. For different Sb doping concentrations, we fabricated nanodevices with Hall bar electrodes to characterize the basic carrier types and densities. The entire set of samples (Table I) shows n-type carrier dominant transport as their Hall resistances are negative values. By introducing more Sb dopants into the ribbon, its sheet resistance increases more than an order of magnitude and the dramatic change in Hall resistance also indicates a much lower carrier density, implying the bulk electron contribution is reduced significantly (Table I). At a high Sb doping concentration (5-7\%), the carrier density reaches 10${}^{13}$cm${}^{-2}$. Considering the carrier density of the surface states from both top and bottom surfaces near the bulk conduction band edge is $\sim$10${}^{13}$cm${}^{-2}$ \cite{Chen2010}, such low carrier density in the Sb-doped nanoribbons suggests that the transport is no longer dominated by large amount of bulk electrons.

\begin{figure}
\centering
\includegraphics[width=80mm]{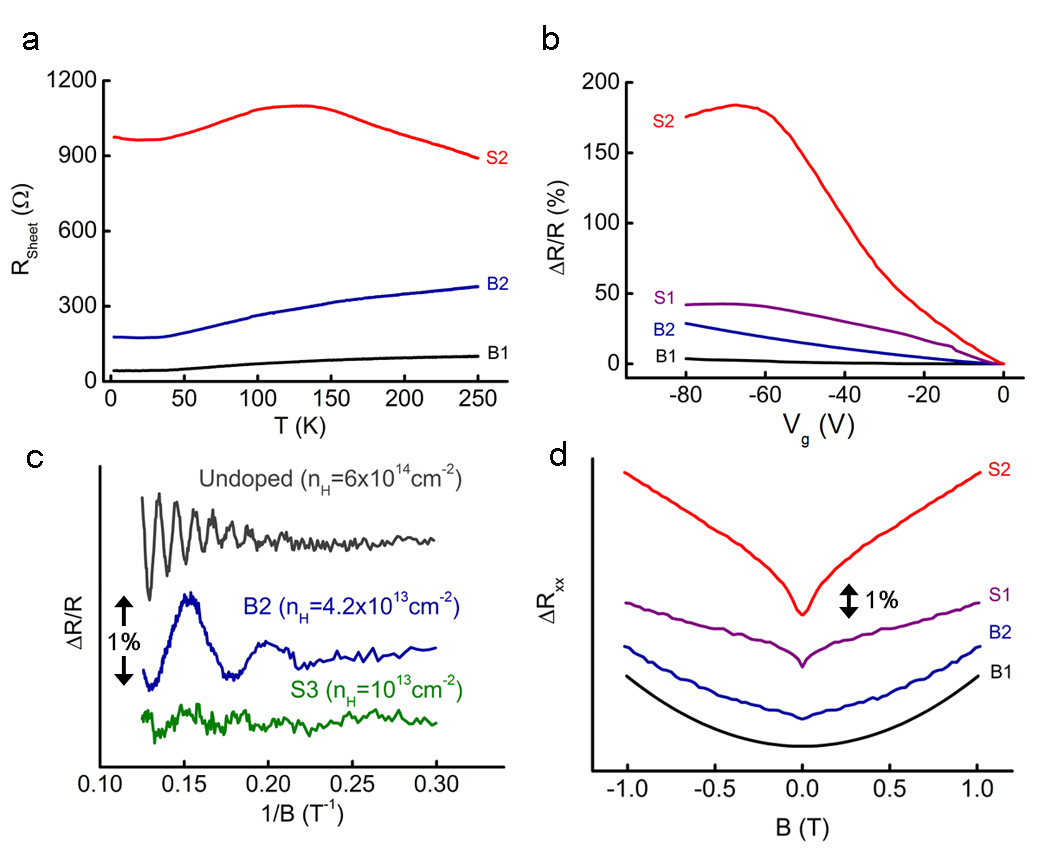}
\caption{\textbf{a}, The resistance profile versus temperature of sample B1, B2, and S2. The most metallic sample (B1) shows a monotonic decrease by cooling. The sample with a small amount of Sb doping (B2) follows a metallic temperature dependence except the small upturn at T \textless 20K. A sample of high Sb concentration (S2) does not follow the typical metallic curve. \textbf{b}, Gating response of the resistance from samples of different doping concentrations (SiO$_2$ 300nm back gate). The resistance of samples B1 and B2 monotonically increases, whereas samples S1 and S2 show maximum peak value near V$_{g}$ = -60V $\sim$ -70V. \textbf{c}, High field magnetoresistance versus inverse magnetic field (1/B). Background curve (either linear or parabolic) is subtracted from original magnetoresistance curve. SdH oscillations from an undoped sample (B${}_{F}$ $\sim$ 93T) corresponds to bulk electron density of $\sim$6\texttimes10${}^{18}$cm${}^{-3}$ and SdH oscillation from sample B2 (B${}_{F}$ $\sim$ 93T) corresponds to bulk electron density of 6\texttimes10${}^{17}$ cm${}^{-3}$. No oscillatory feature is observed from sample S1, S2, and S3 within the field limit (8T). The curves are displayed with an offset for visual clarity. \textbf{d}, Magnetoresistance near zero magnetic field, showing clear feature of weak anti-localization from samples of high Sb concentration (S1, S2). Each curve is normalized by zero field resistance and displayed with an offset for visual clarity.}
\end{figure}

Additional electronic transport studies confirm that the bulk electron contribution is reduced significantly by the Sb doping. In Figure 2a, temperature dependent resistances from low Sb concentration samples (B1, B2) follow typical metallic behavior. In contrast, for the sample of high Sb concentration (S2), the resistance starts to increase and saturates at low temperature. An increase in resistance upon reducing temperature is likely due to freezing out of the bulk carriers. Moreover, electrostatic gating experiments in field effect transistor devices manifest drastic difference between low Sb concentration samples and high Sb concentration samples (Fig. 2b). Low Sb-doped samples (B1, B2) show weak gating dependence by a bottom gate, which is reasonable as the ribbon thicknesses (\textgreater100nm) is much larger than the depletion layer thickness ($\sim$10nm) with still relatively high carrier concentration. However, the gating response becomes larger with increasing Sb doping concentration. The sample with the highest Sb doping (S2) exhibits the largest increase of its resistance, or the most significant decrease in bulk carrier concentration. The conductance of this ribbon decreases by more than half by electrostatic manipulation with the bottom gate in spite of its large thickness (120nm), which strongly suggests the increase of the depletion layer thickness ($\sim$35nm) and the significant suppression of the bulk transport contribution.

Magnetotransport data clearly shows the emergence of the surface state and suppression of bulk electrons by Sb doping as well. In the high field magnetoresistance (MR), quantum oscillations (SdH oscillations) are observed in samples of different Sb concentrations (Fig. 2c). The oscillation does not depend on the gate voltage (V$_{g}$), which suggests that the oscillation originates from bulk electrons. Without Sb doping, the bulk SdH oscillation periodicity in an inverse magnetic field is small. It indicates the cross section of the Fermi surface is large, consistent with high bulk carrier concentration. As Sb concentration increases, the oscillation period gets larger and disappears at the highest doping level. That is, the bulk electron pocket crossing the Fermi level eventually becomes too small so that the oscillation period is too large to be detected in our magnetic field range (8T). The small cross section of the bulk Fermi surface means the contribution of bulk electrons is greatly reduced. In the low magnetic field, weak anti-localization (WAL), the quantum correction which emerges strongly in the spin-orbit coupled surface state\cite{Chen2010}, is absent in the MR trace of the undoped sample (Fig. 2d, B1) as the large bulk electron contribution masks surface state transport. On the contrary, higher Sb concentration samples (Fig. 2d, S1, S2) manifest a sharp cusp near zero magnetic field, which is a characteristic feature of WAL. Owing to the reduced bulk electron contribution by Sb doping, all the transport measurements including temperature dependence, field effect gating, and magnetotransport consistently indicate significant contribution from surface state transport.

Even though the 5-7$\%$ Sb doping in the aforementioned experiments can effectively reduce the carrier concentration to $\sim$10${}^{13}$cm${}^{-2}$ and lower the Fermi level close to the conduction band edge, nanoribbons are still exposed to extrinsic contaminations during device fabrication, which have the issue of the additional increase of the bulk carrier concentration, as reported by previous studies\cite{Analytis2010, Kong2011ACS}. Therefore, we hypothesize that the Sb-doped Bi$_2$Se$_3$ should have much lower carrier concentration if protected from the ambient environment. Here we deposit an insulating Zinc Oxide (ZnO) layer on top of the nanoribbons to protect samples from external contamination. The sputtered ZnO layer covers the entire surface of the nanoribbons, which would prevent degradation and extrinsic doping that might happen during the standard fabrication process.

\begin{figure}
\centering
\includegraphics[width=75mm]{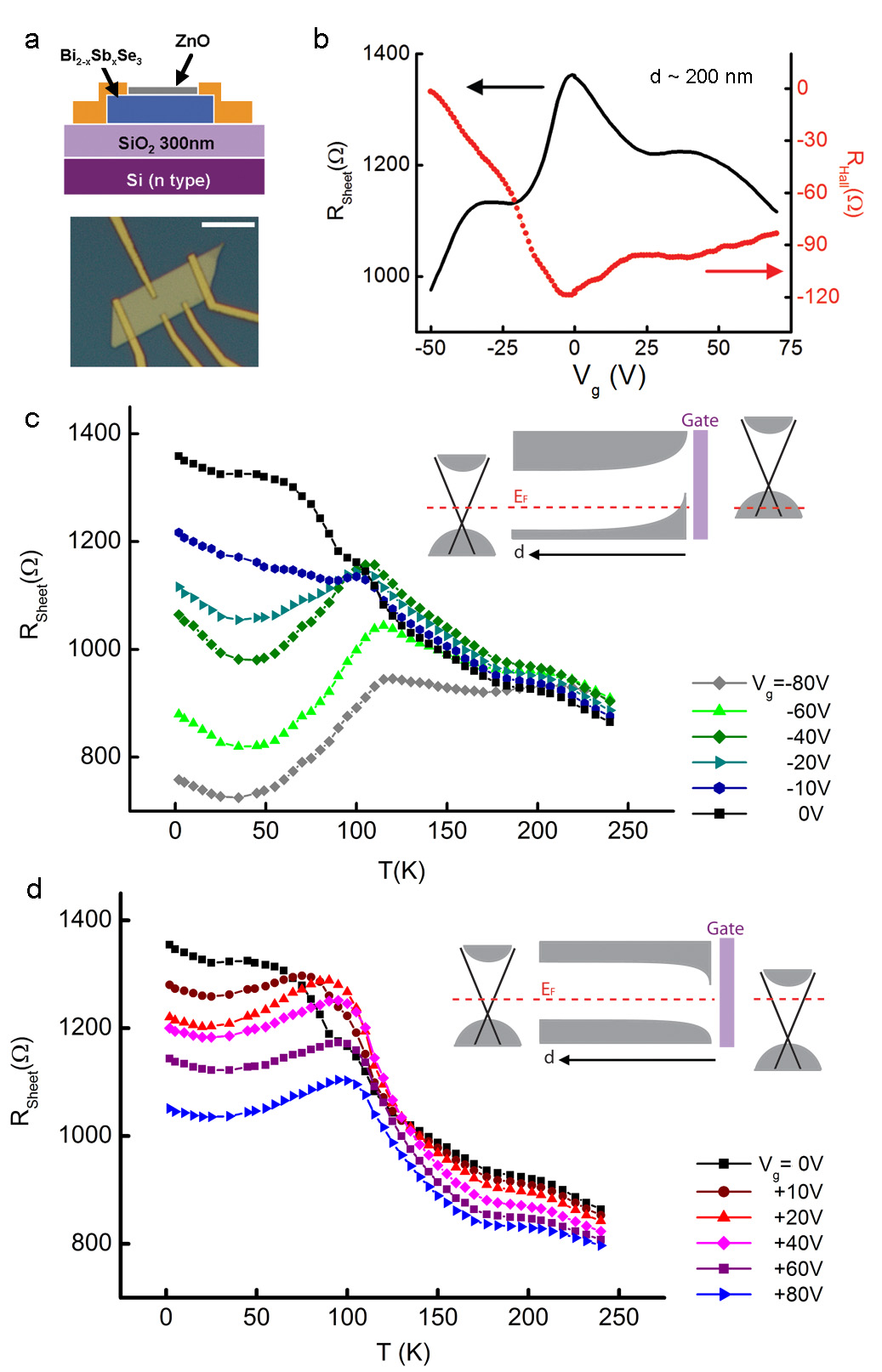}

\caption{\textbf{a}, Device schematic (top) and optical image (bottom) of a thick sample ($\sim$200nm). Scale bar is 5$\mu$m. \textbf{b}, Gate voltage dependence of longitudinal sheet resistance (R${}_{sheet}$) and Hall resistance (R${}_{xy}$) measured at low field (B \textless 2T) at T=2K. Hall resistance increases in both direction of V$_{g}$, implying that the Fermi level of the bottom surface is near the Dirac point at zero gating voltage. The anomalous kink in the longitudinal resistance curve is not well understood, and it is not reproducible in other samples. \textbf{c}, Temperature-dependent resistance curve at different V$_{g}$ (negative). The curve changes significantly as more holes are added, reflecting the conventional metallic temperature dependence of induced carriers. A band diagram (inset) shows band bending at the bottom surface induced by gating. \textbf{d}, Temperature dependent resistance curve at different gate voltages (positive). In a band diagram (inset), the Fermi level does not cross any bulk band by positive gating, explaining the qualitatively similar temperature curves over the wide range of gating voltage.}
\end{figure}

Here we fabricate a bottom-gate device using Sb-doped and ZnO protected Bi$_2$Se$_3$ nanoribbon with a large thickness ($\sim$200nm), as shown in Fig. 3a, The gating is effective in the range of $\sim$30nm for this carrier concentration, and the thickness of the ribbons is much larger than this effective gating layer thickness, so that the electrostatic bottom gate can only affect the bottom surface state. We have studied gate-dependence and temperature-dependence of resistance, which can examine the nature of carriers induced by gating voltage. Owing to the Sb doping and ZnO protective layer as discussed above, we found that the Fermi level of the bottom surface in this particular device is close to the Dirac point. In Figure 3b, its resistance decreases by either direction of gating voltage, and Hall resistance increases as more electronic states on the bottom surface (p-type carriers for negative V$_{g}$ and n-type carriers for positive V$_{g}$) are populated. At zero V$_{g}$, resistance increases with decreasing temperature, and saturates at low temperature, similar to the case of Sb-doped bulk crystal\cite{Analytis2010}. The increase of resistance at high temperature (T \textgreater 100K) can be understood as the freeze out of the thermally populated bulk electronic states. As p-type carriers are induced by the negative V$_{g}$ (Fig. 3c), there is a noticeable change in the temperature curve - resistance starts to drop down by cooling. As we have observed this metallic temperature dependence when the Fermi level is crossing the bulk band, we conclude that it becomes a mixed state with bulk carriers as increasing p-type carriers. On the other hand, the temperature curve does not change qualitatively by inducing more n-type carriers (positive V$_{g}$, Fig. 3d).

The asymmetric trends in the temperature dependent transport reflect the characteristic band structure near the Dirac point in Bi$_2$Se$_3$ topological insulators. From ARPES studies\cite{Xia2009, Analytis2010PRB}, the Dirac point of surface states is just above the bulk valance band edge, while it is relatively far apart from the bulk conduction band edge ($\sim$0.2eV). Therefore, within our V$_{g}$ range (+80V), positive gating only creates more electrons from the surface band but not from the bulk band, which will not change the overall shape of the temperature curve. In contrast, bulk electronic states are easily populated from the valence band by negative gating. Now, the ribbon is in the mixed state, evolving from surface dominant transport to bulk dominant transport by adding more bulk hole carriers by gating, showing the dramatic change in the temperature-dependent resistance curves. This temperature dependence study clearly shows that we can independently tune the Fermi level of one surface near the Dirac point. This is an interesting system to controllably build either a topological insulator junction of dual types of carriers or a single surface junction to study novel proximity effects in the future\cite{Seradjeh2009, Fu2008}.

\begin{figure}
\centering
\includegraphics[width=75mm]{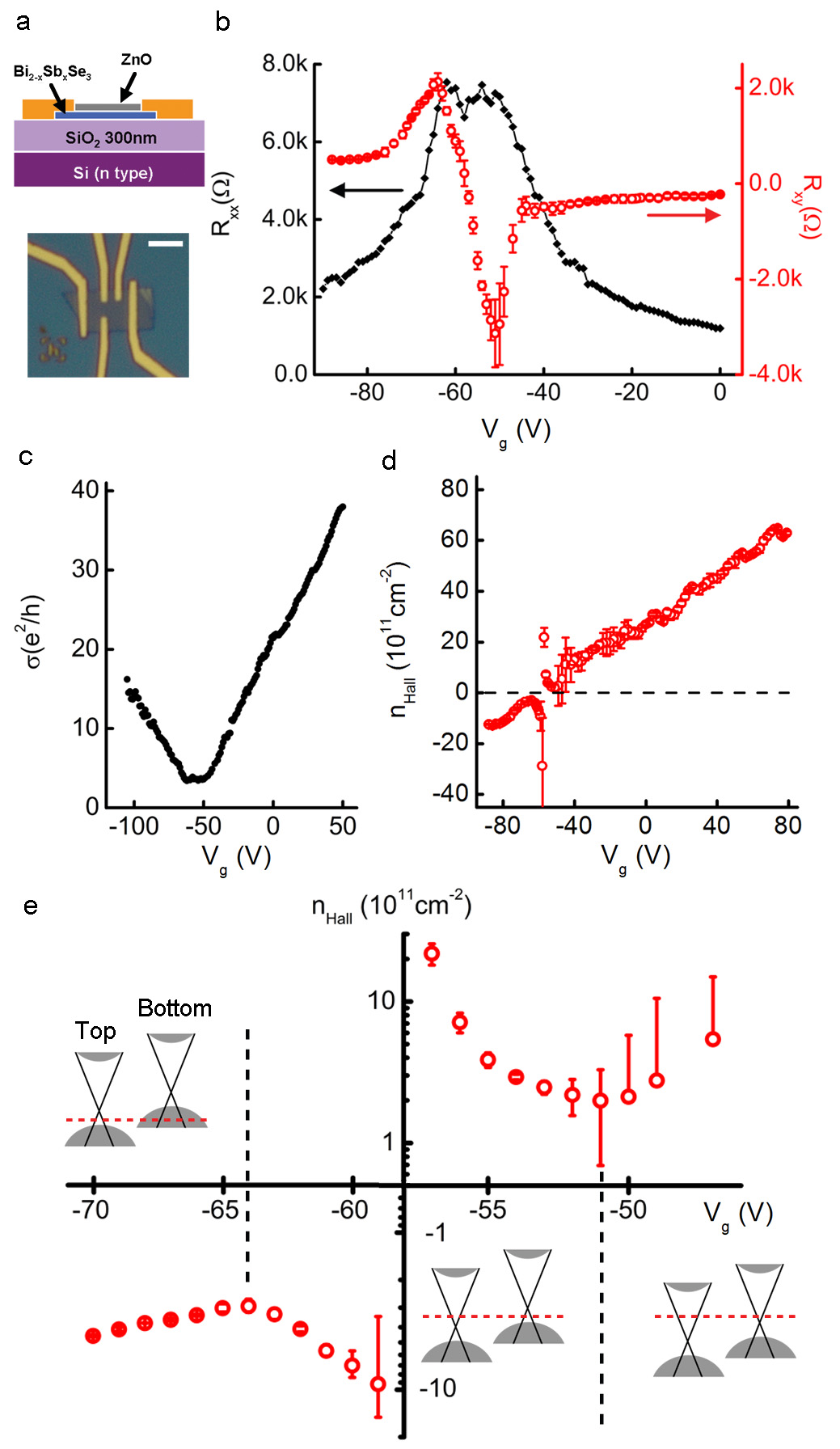}
\caption{\textbf{a}, Device schematic (top) and optical image (bottom) of thin ribbon sample ($\sim$6nm). Scale bar indicates 3$\mu$m. \textbf{b}, Gate voltage dependence of longitudinal resistance (R${}_{xx}$) and Hall resistance (R${}_{xy}$) measured at low magnetic field (B \textless 2T), at T=2K. Hall resistance decreases as n-type carriers are depleted (V$_{g}$ \textgreater -50V); increases as induced p-type carriers from the bottom surface form a mixed state with decreasing n-type carriers (-65V \textless V$_{g}$  \textless -50V); and decreases again as the entire sample becomes a hole conductor (V$_{g}$  \textless -65V).
\textbf{c}, Conductance ($\sigma$) versus V$_{g}$ curve is linear, except near the minimum conductance. \textbf{d}, Electron density (n$_{Hall}$) plot as a function of V$_{g}$. It depends on V$_{g}$ linearly in the wide range of voltage (-90V to +80V). \textbf{e}, Semi-log scale plot of carrier density near its charge neutrality point (Dirac point). Band diagrams of top and bottom surfaces (inset) for samples with pure p-type conduction (left), mixed conduction (middle), and pure n-type conduction (right).}
\end{figure}

In addition to the manipulation of one surface in the thick ribbons, we eventually want to control the Fermi level of the both top and bottom surfaces to achieve ultralow carrier concentration near the Dirac point. To shift the Fermi level of the entire ribbon together by electrostatic gating, the nanoribbons need to be further thinned down. We etched a thick, Sb-doped ribbon ($\sim$100nm) by Argon plasma in a sputtering machine and in-situ deposited a 15nm thick ZnO protection layer. In Fig. 4a, the ribbon is semitransparent with a thickness of 6nm (confirmed by AFM), more than 30 times thinner than the previous device (Fig. 3). From this 6nm-thin device, we can achieve very low electron concentration of 2.8\texttimes10${}^{12}$cm${}^{-2}$ at zero V$_{g}$ (Fig. 4b and 4d). Since this concentration value is more than three times lower than that ($\sim$10${}^{13}$cm${}^{-2}$) of the surface states from the both top and bottom surfaces of which Fermi level is near the bulk conduction band edge, we believe that the Fermi level of the both top and bottom surfaces is completely within the bulk band gap. In other words, the Fermi level crosses only the surface Dirac cone above the Dirac point.

In this ZnO-protected nanoribbon device already with very low carrier concentration, its small thickness of 6nm makes it possible to shift the Fermi level of the entire ribbon by electrostatic gating across the Dirac point. When V$_{g}$ is changed from positive to negative bias, its longitudinal resistance (R${}_{xx}$) first increases and reaches a large peak value ($\sim$7k$\Omega $) at V$_{g}$ of -50V, and then decreases when further lowering the gating voltages below V$_{g}$ of -65V (black curve, Fig. 4b). The Hall resistance (red curve, Fig. 4b) changes significantly too, as its magnitude increases by more than 10 times and switches its sign also at V$_{g}$ of -60V. These results clearly demonstrate the ambipolar field effect and suggest that the entire sample is converted from n-type to p-type. The sample conductance depends on gating voltage linearly, except at the plateau of minimum conductance ($\sim$3.6G$_0$) (Fig. 4c). The absence of zero conductance region during the ambipolar transition is due to the gapless surface states. The origin of the conductance plateau is not definitive yet, which could be explained as suggested in previous theoretical studies\cite{Liu2010,Linder2009,Culcer2010}. The carrier density obtained by Hall resistance is plotted in Figure 4d showing its linear dependence on V$_{g}$ as well. The sample stays pure n-type until V$_{g}$ = -50V, and then switches to a mixed carrier state (-65V \textless V$_{g}$ \textless -50V) and p-type (V$_{g}$ \textless -65V) subsequently (Fig. 4e). Considering the order of low carrier density (\textless 10$^{12}$ cm${}^{-2}$), now both surfaces are very close to the Dirac point in this range of V$_{g}$. In fact, the minimum carrier density obtained from Hall resistance is about 2\texttimes10${}^{11}$cm${}^{-2}$, which is 10 times lower than the concentration at zero V$_{g}$. This low carrier density value is also comparable with that of initial single layer graphene devices\cite{Zhang2005, Novoselov2005}.

The Sb doping of topological insulator nanoribbons and the protective oxide layer effectively reduce bulk carriers originating from both intrinsic and extrinsic material defects. Consequently, we are able to carry out electronic transport measurement to demonstrate surface state dominated transport as a result of fairly low bulk carrier concentration. In the thick sample, single surface can be individually tuned within the wide range of the Fermi level. In the thin sample, both top and bottom surfaces are easily tunable across the Dirac point by electrostatic gating. With minimal contribution from residual bulk carriers, Sb-doped Bi$_2$Se$_3$ nanoribbons will offer a great opportunity to test various topological insulator phenomena proposed by theory, as well as be the optimal material for future applications.

\subsection{Acknowledgments}

\begin{acknowledgments}
We thank K. Lai and J. R. Williams for the helpful discussions, and B. Weil for the help in the manuscript preparation. Y.C. acknowledges the supports from the Keck Foundation, DARPA MESO project (N66001-11-1-4105), and King Abdullah University of Science and Technology (KAUST) Investigator Award (No. KUS-l1-001-12). 
\end{acknowledgments}


\end{document}